\documentclass{kluwer}
\usepackage{epsfig}
\def\be{\begin{equation}}
\def\ee{\end{equation}}

\def\jcd{Christensen-Dalsgaard}

\def\ga{{\leavevmode\kern0.3em\raise.3ex\hbox{$>$}
\kern-0.8em\lower.7ex \hbox{$\sim$}\kern0.3em}}
\def\la{{\leavevmode\kern0.3em\raise.3ex\hbox{$<$}
\kern-0.8em\lower.7ex \hbox{$\sim$}\kern0.3em}}

\begin{document}
\begin{article}
\begin{opening}

\title{SOLAR CYCLE VARIATION IN SOLAR F-MODE FREQUENCIES AND RADIUS}
\runningtitle{Solar cycle variation in f-mode frequencies}
\runningauthor{H. M. Antia et al.}

\author{H. M. \surname{Antia}}
\institute{Tata Institute of Fundamental Research,
Homi Bhabha Road, Mumbai 400005, India}
\author{Sarbani \surname{Basu}}
\institute{Institute for Advanced Study,
Olden Lane, Princeton NJ 08540, U. S. A., and 
Astronomy Deparment, Yale University, P.O. Box 208101 New Haven,
CT 06520-8101 USA}
\author{J. \surname{Pintar} and B. \surname{Pohl}\thanks{present
address: US Naval Observatories, Washington, DC 20392-5420, U. S. A.}}
\institute{National Solar Observatory, PO Box 26732, Tucson
AZ 85726-6732, U. S. A.}
\date{\today}

\begin{abstract}
Using data from the Global Oscillation Network Group (GONG)
covering the period from 1995 to 1998, we study the change with solar
activity in 
solar f-mode frequencies. The results are compared with
similar changes detected from the Michelson Doppler Imager (MDI) data. 
We find variations in f-mode
frequencies which are correlated with solar activity indices. If these
changes are due to variation in solar radius then the implications are that
the solar radius decreases by about 5 km from minimum to maximum activity.
\end{abstract}

\end{opening}
\section{Introduction}
The fundamental mode or f-mode of solar oscillations are believed to
be surface gravity modes whose frequencies are essentially independent
of the stratification in the solar interior. A large part of the
difference between the observed f-mode frequencies at intermediate
degree and those in a standard solar model has been interpreted as being
caused by the incorrect radius of the solar models used (Schou et al.~1997;
Antia~1998; Brown and \jcd~1998; Tripathy and Antia 1999). The
frequencies of these modes can thus be used to measure the solar radius. 

There have been many reports about possible variation of solar radius
with time (Delache, Laclare and Sadsaoud~1985; Wittmann, Alge and
Bianda~1993; Fiala, Dunham and Sofia~1994; Laclare et al.~1996; Noeel
1997). The reported change in measured angular semi-diameter varies from
$0.1''$ to $1''$, which implies a change of 70 to 700 km in radius.
However, there is no agreement among observers about these variations.
It would thus be interesting to look for corresponding variations in the
f-mode frequencies. The reported variations in solar radius should
change the f-mode frequencies by $0.1$ to $1\;\mu$Hz, which is much
larger than the error estimates in these frequencies. Considering 
how small the
estimated errors in f-mode frequencies are, it is possible to measure changes
in the solar radius as small as a few kilometers over the solar cycle. 
Some variation in f-mode frequencies have been reported in the MDI data
by Dziembowski et al.~(1998) who find that the solar radius reached a
minimum around the minimum activity period and was larger by about 5 km,
6 months before and after the minimum. If this variation is correlated with
solar activity then we would expect much larger change between minimum
and maximum activity. 

In this work we look for possible variations in the solar radius using
GONG data --- which extend over longer time period than the MDI data
used in previous studies. We also look for possible correlations between
change in radius and solar activity. In Section~2 we outline the
technique used in calculating the frequencies and examine possible
systematic errors in computed frequencies. Section~3 gives the results
while in Section~4 we summarize the conclusions from our study.

\section{The technique}
Since the f-modes have very low power they are barely visible in the
spectra for modes of a given degree $\ell$ and azimuthal order $m$ that 
the GONG project fits to determine solar oscillation frequencies. As a
result, it is difficult to determine the frequencies of these modes
reliably from these spectra. To improve statistics we use the 
rotation-corrected, $m$-averaged, power spectra from GONG data 
(Pohl and Anderson
1998) to calculate the mean f-mode frequencies. Because summing
over the $2\ell+1$ spectra for a  mode of degree $\ell$ makes the peaks 
better defined,  it is possible to fit them reliably. Each of these spectra
was obtained from 3 GONG months (108 days) of data and has a
frequency resolution of
0.107 $\mu$Hz. These spectra extend up to $\ell=200$ and frequency 
$\nu=2083\;\mu$Hz. The f-mode peaks are clearly visible for
$\ell\ga100$. We have also used two spectra covering only 1 GONG month,
which gives a lower frequency resolution of 0.321 $\mu$Hz.
For these spectra the errors in computed frequency would be larger
as compared to the 3 month spectra.

We fit each mode separately using a maximum likelihood technique
(Anderson, Duvall and Jefferies~1991) with a symmetric Lorentzian
profile for the peaks. The fit is performed over a region extending to
about 20 $\mu$Hz on either side of the peak and includes leaks from modes
of degree
$\ell-3$ to $\ell+3$, $\ell$ being the degree of the target mode.
 It is found that peaks arising from the $\ell+1$
leak are split into two because of leaks from peaks of $m\pm1$. Similarly, the
$\ell+2$ leak is split into three because of leaks from peaks of  $m,m\pm2$.
Although leaks from $\ell-1$ and $\ell-2$ are also split in a similar
manner, the power in these peaks is much smaller. Thus we have
explicitly fitted the $m$-leaks only from the $\ell+1$ and $\ell+2$ peaks.
An example of the fit is shown in Fig.~1. 

\begin{figure}
\centerline{\epsfig{file=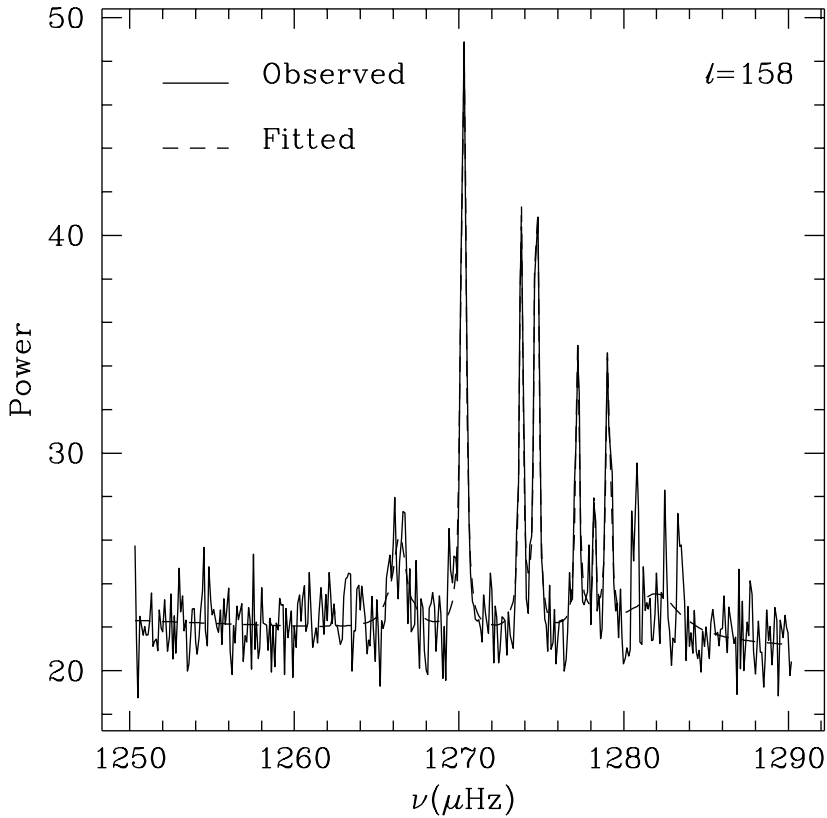,width=11.0 true cm}}
\caption{
Fit to the $\ell=158$ f-mode spectra for months 33--35. Power is in
arbitrary units.}
\end{figure}

For $\nu<1170\;\mu$Hz, the modes are difficult to fit reliably since the
width is smaller than the resolution limit of the spectra and the power
is relatively low. As a result, in this work we only include higher
frequency modes.

The use of $m$-averaged spectra may introduce some systematic errors in
determining the frequencies. To check for these we compare the
frequencies of p-modes fitted from $m$-averaged spectra with those
obtained by the GONG project using the individual $m$ spectra. The GONG
project has computed the mean frequency for each $n,\ell$ multiplet by
fitting the frequencies for individual values of $m$ to polynomials in
$m$. This mean frequency can be compared with the frequency computed
from the $m$-averaged spectra. The results are shown in Fig.~2. It may be
noted that f-mode has not been fitted in the individual $m$ spectra
and hence these are not included. It can be seen that the
frequency difference is quite small being of the order of the estimated
errors in the fitted frequencies. Systematic difference between these
two sets of frequencies is $\la10$ nHz. 

\begin{figure}
\centerline{\epsfig{file=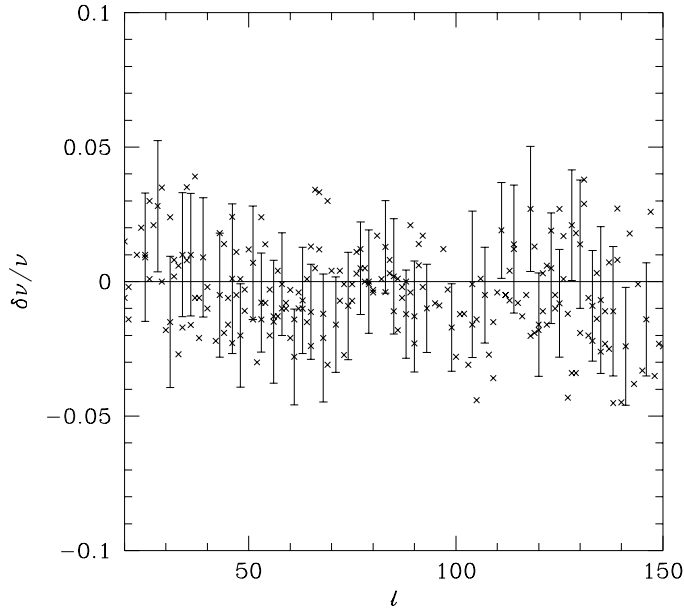,width=9.0 true cm}}
\caption{
Difference in frequencies for months 33--35 between the values obtained
from the $m$-averaged spectra and those computed from the individual $m$-spectra.
For clarity only a few error-bars are shown.}
\end{figure}

Another possible source of systematic errors is the use of incorrect even-order
splitting coefficients while constructing the $m$-averaged spectra. The
GONG spectra were obtained by setting these coefficients to zero. A
non-zero value of these even-order coefficients would introduce some
systematic shift in frequencies. To estimate the effect of this we
constructed some $m$-averaged spectra with either $a_2=-0.2$ nHz or
$a_4=0.1$ nHz. The resulting frequency shifts with $a_2=-0.2$ nHz
are shown in Fig.~3. Similar shifts are found for non-zero $a_4$. Once
again the frequency shifts due to non-zero $a_2$ or $a_4$ are quite
small, being of the order of 10 nHz. Furthermore, the frequency shift
for the f-mode is smaller than those for p-mode with the same $\ell$.
Although these even coefficients are known to change with solar cycle,
the variation is not very large at the low frequencies which we are
considering in this work. Thus the shifts shown in Fig.~3 can be
considered to be  conservative upper limits to errors that may be expected
due to non-zero even coefficients. From Fig.~2 it can be seen that
during a period where even coefficients are fairly large the difference
in frequencies computed from individual $m$-spectra and the $m$-averaged
spectra is quite small --- much smaller than the differences seen in
Fig.~3. 

\begin{figure}
\centerline{\epsfig{file=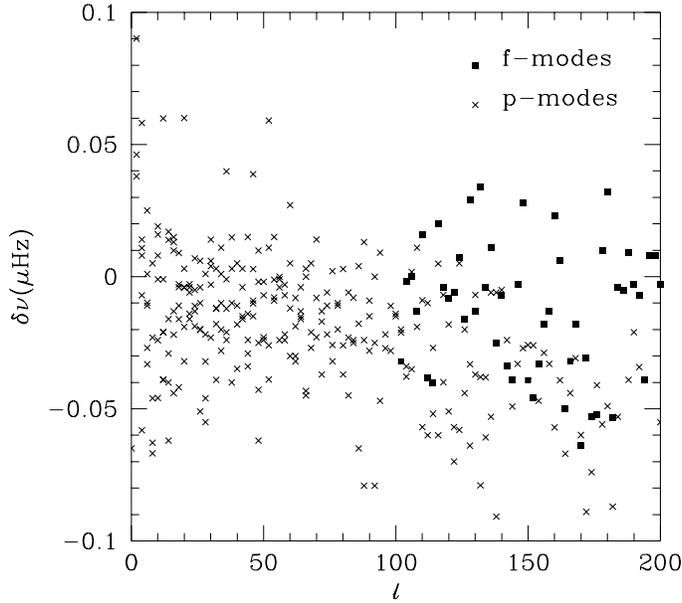,width=9.0 true cm}}
\caption{
Shift in frequencies for months 21--23 arising due to non-zero $a_2$
}
\end{figure}

Considering possible systematic errors, it would be desirable to
restrict ourselves to low frequency range where these errors are not
significant. In the frequency range ($\nu<1440\;\mu$Hz) considered in
this work the systematic errors from these effects are expected to be
about 10 nHz, and any time variation in these systematic effects would
be smaller. Having established the reliability of frequencies from
$m$-averaged spectra we calculate the frequencies from 7 different spectra
during 1995 to 1998. These spectra are listed in Table~1. This table
also lists the solar activity as measured by the mean daily sunspot
number, $R_I$ from the Solar Geophysical Data web page
 of the US National Geophysical
Data Center (http://www.ngdc.noaa.gov/stp/stp.html).

\begin{table}
\caption[]{List of GONG spectra used}
\begin{tabular}[]{cccc}
\hline
\noalign{\smallskip}
GONG months&Starting date&Ending date&\hbox{Mean daily sunspot no.}\cr
\noalign{\smallskip}
\hline
\noalign{\smallskip}
3 -- 5&1995-07-18&1995-11-02&$14.5\pm1.3$\cr
\phantom{0}8 -- 10&1996-01-14&1996-04-30&$\phantom{0}6.0\pm0.6$\cr
21 -- 23&1997-04-26&1997-08-11&$14.7\pm1.1$\cr
27 -- 29&1997-11-28&1998-03-15&$39.5\pm1.7$\cr
33 -- 35&1998-07-02&1998-10-17&$79.5\pm2.5$\cr
36&1998-10-18&1998-11-22&$62.4\pm4.1$\cr
37&1998-11-23&1998-12-28&$82.6\pm4.1$\cr
\noalign{\smallskip}
\hline
\end{tabular}
\end{table}

\section{Results}
We have fitted each of the 7 spectra listed in Table~1 to calculate the
mean frequencies of the f modes. Fig.~4 shows the relative difference
between the observed frequencies for Months 33--35 and Months 8--10.
It can be seen that to a first
approximation the relative frequency difference is independent of
frequency and hence this frequency difference can be interpreted
as arising from a change in the solar radius. Of course, suitable
combination of other effects like magnetic field, convection etc.
may also yield frequency differences which are independent of frequency
but that will almost certainly require some fine-tuning of parameters,
while change in solar radius would be a simpler explanation.

\begin{figure}
\centerline{\epsfig{file=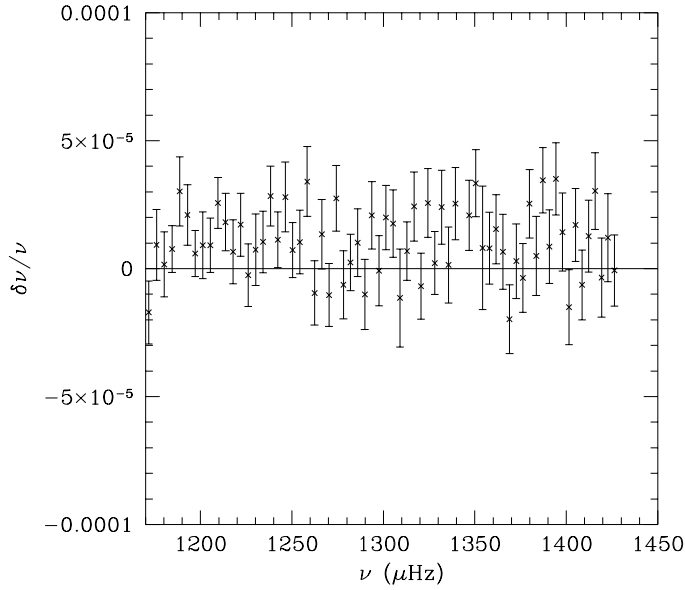,width=9.0 true cm}}
\caption{Relative frequency differences between
f-mode frequencies from the
spectra for months 33--35 and months 8--10.}
\end{figure}

To estimate the possible change in radius we take an average over
the relative frequency difference for all 56 f-modes that were successfully
fitted in all spectra:
\be
\langle \delta\nu\rangle={\sum_\ell {\delta \nu_\ell\over
\sigma_\ell^2}\over
\sum_\ell{1\over\sigma_\ell^2}}
\label{eq:delnu}
\ee
The differences were taken with respect to calculated frequencies
for a standard solar model, which is
a static model,
constructed with OPAL opacities (Iglesias \& Rogers 1996) and
low temperature opacities
from Kurucz (1991).
The OPAL equation of state (Rogers, Swenson \& Iglesias 1996) 
was used
to construct the model and convective flux was calculated using the 
the formulation of Canuto and Mazzitelli (1991).
The radius of the model is 695.78 Mm (Antia 1998).
The results are shown in Fig.~5, which also shows
the mean daily sunspot number, $R_I$, for each of these periods.
The change in frequencies appears to be correlated with the
sunspot number.

\begin{figure}
\centerline{\epsfig{file=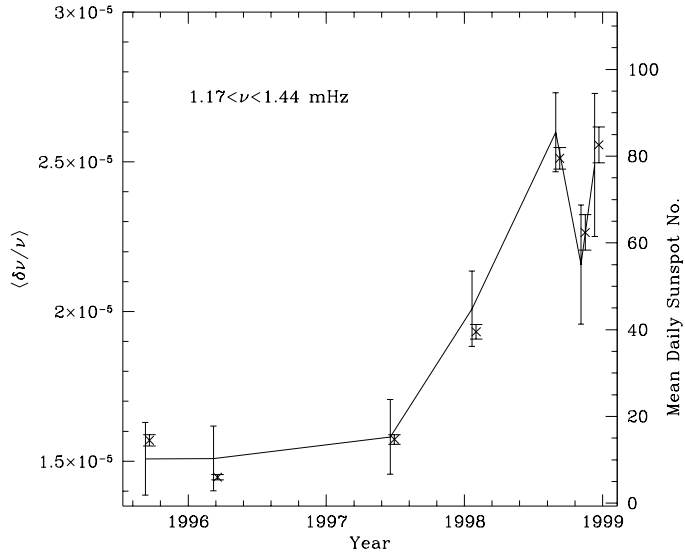,width=9.0 true cm}}
\caption{The averaged frequency difference
for f-modes between the Sun and a standard solar model with radius
$R=695.78$
Mm as a function of time is shown by the continuous line.
The mean daily sunspot number
$R_I$ is also shown by points with scale shown on the right side.
For clarity these points have been slightly shifted along the x-axis.}
\end{figure}

If the change in frequencies is entirely due to change in radius then
the average frequency difference is related
to change in radius (Antia 1998) by
\be
\langle{\delta\nu\over\nu}\rangle =-{3\over2}{\delta R\over R}
\label{eq:reldelnu}
\ee
Thus a relative shift in frequencies by $10^{-5}$ corresponds to a
change in radius by $0.67\times10^{-5}R_\odot\approx4.7$ km. It may be
argued that the change of frequency may be due to some other effect,
e.g., magnetic field. Since the relative frequency difference between
the frequencies at two different epochs is almost independent of
frequency in the range of degrees considered here, a change in
radius is the simplest explanation for the frequency shift.
Of course, the magnetic field is directly correlated to activity indices,
but the direct effect of 
magnetic field, is likely to yield frequency differences
that will increase rapidly with frequency since the higher frequency modes
are located closer to solar surface where these effects are likely to be
more important (Campbell and Roberts 1989). 
It may be noted that a reduction in solar radius by 4.7 km in 2 years
will generate about 100 times the solar luminosity through release of
gravitational potential energy, if the contraction in homologous.
This indicates that the change in radius is a superficial effect
confined to outer layers where the f-modes are located.

However, the radius appears to decrease with increasing activity. This
is not what is seen by Dziembowski et al.~(1998). The magnitude
of change found in GONG data during the period covered by their study is
much smaller than what they have reported. 
The apparent discrepancy is
most likely due to a different range of degrees used in their
study. They have probably used modes up to $\ell=300$ and these high
wavenumber modes are known to deviate significantly from the simple
dispersion relation (Duvall, Kosovichev and Murawski~1998; Antia and
Basu 1999). As a result, one can expect misleading results if average is
taken over all modes, since it is possible that the deviation of the
modes from the normal dispersion relation has some solar cycle dependence.
A part
of the deviation is known to be due to the use of symmetric profiles in
calculating the f-mode frequencies (Antia and Basu 1999) and it is quite
possible that the asymmetry changes with activity. Thus it would be
better to confine our attention to modes with $\ell<200$, where these
effects are minimal. The results shown in Fig.~2 of Dziembowski et
al.~(1998) imply a decrease in frequency by $10^{-5}$ between 1996.8 and
1997.2. If this change is correlated with sunspot number then we would
expect a decrease by at least $5\times10^{-5}$ between 1996 and 1998.6
in our results. Certainly no such shift is seen in our results. This
shift is also much larger than the expected systematic errors in our
results. 

\begin{figure}
\centerline{\epsfig{file=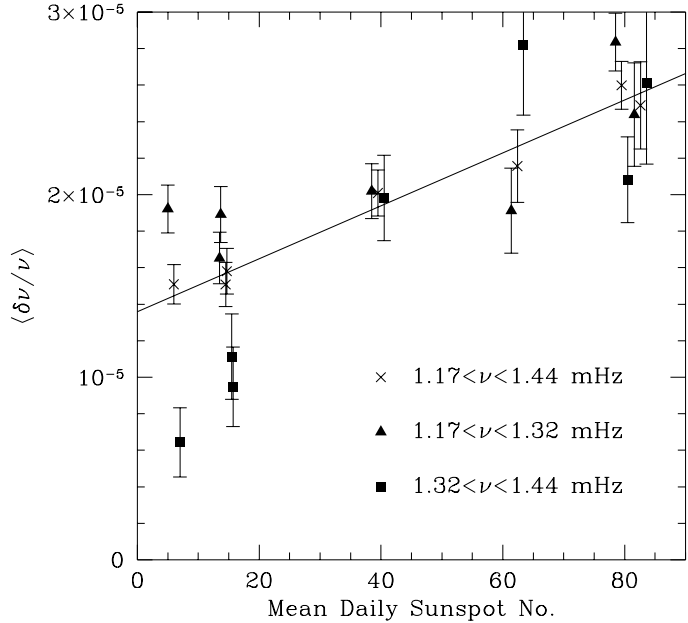,width=9.0 true cm}}
\caption{The averaged frequency difference
for f-modes between the Sun and a standard solar model with radius
$R=695.78$
Mm are plotted against the mean sunspot number during the period of
observation. The average over three different frequency ranges
as marked in the figure are compared with each other. The line marks the best linear
fit for the full frequency range. The triangles and squares
are slightly shifted along $x$-axis for clarity.}
\end{figure}

In order to check the stability of our results we have repeated the
analysis by restricting the frequencies to $1.17<\nu<1.32$ mHz
and $1.32<\nu<1.44$ mHz and the
results are shown in Fig.~6.
It is clear that the results are not particularly sensitive to
the range of frequencies used. However, the correlation with solar
activity is markedly weaker in the higher frequency range.
From the relative frequency differences between the observed frequencies
and those in a standard solar model it is known (Tripathy \& Antia 1999)
that there is a distinct tendency of relative frequency
differences decreasing with frequency in this range. It is likely that
this trend is due to some other effect like magnetic field which varies
with solar cycle. In that case it will interfere with the signal due to
change in radius. It is quite possible that at even higher frequencies
used in MDI study, the correlation is further weakened because there is
an even stronger decrease in the relative frequency differences.

\section{Conclusions}
We have calculated the mean f-mode frequencies from the $m$-averaged
spectra from GONG data covering 108 days. The use of $m$-averaged spectra
enable us to determine the frequencies reliably. The relative frequency
difference between observed frequencies and those calculated for a
standard solar model are roughly independent of $\ell$ in the range that
was used in this study ($100<\ell<200$). Hence it is likely that this
difference is due to difference in solar radius, as other agents are
likely to introduce a steep frequency dependence in relative frequency
differences.
The mean shift in f-mode
frequencies appears to be correlated with solar activity indices like
the mean daily sunspot number $R_I$. 

If this frequency shift is due to change in solar radius then the radius
decreases as activity increases. Further, the change in the solar radius
during the solar cycle is approximately 5 km. This is one or two orders
of magnitude smaller than the changes reported earlier (Delache, Laclare
and Sadsaoud~1985; Wittmann, Alge and Bianda~1993; Fiala, Dunham and
Sofia~1994; Laclare et al.~1996; Noeel 1997). 

These results on the change in solar radius with solar cycle are
much less than what has been deduced using MDI data (Dziembowski et
al.~1998) and we believe the discrepancy is due to their using higher
degree modes whose frequencies can be affected by a variety of other
agents, since these modes are localized closer to the solar surface.

\acknowledgements
This work utilizes data obtained by the Global Oscillation Network Group
(GONG) project, managed by the National Solar Observatory, a Division of
the National Optical Astronomy Observatories, which is operated by AURA,
Inc. under a cooperative agreement with the National Science Foundation.
The data were acquired by instruments operated by the Big Bear Solar
Observatory, High Altitude Observatory, Learmonth Solar Observatory,
Udaipur Solar Observatory, Instituto de Astrofisico de Canarias, and
Cerro Tololo Inter-American Observatory.

\end{article}
\end{document}